\documentclass[11pt,twoside]{article}
\usepackage{cspm2015}

\aspSuppressVolSlug
\resetcounters

\def\arcsec{\hbox{$^{\prime\prime}$}}
\begin{document}

\title{Multi-wavelength observations of vortex-like flows in the photosphere using ground-based and space-borne telescopes}

\author{J.~Palacios,$^1$ S.~Vargas Dom{\'{\i}}nguez,$^2$ L.~A.~Balmaceda,$^{3,4}$ I.~Cabello$^5$ and V.~Domingo$^6$
\affil{$^1$ Universidad de Alcal{\'a}, Alcal{\'a} de Henares, Madrid, Spain; \email{judith.palacios@uah.es}}
\affil{$^2$ Universidad Nacional de Colombia, Bogot{\'a}, Colombia; \email{svargasd@unal.edu.co}}
\affil{$^3$ICATE-CONICET, San Juan, Argentina; \email{lbalmaceda@icate-conicet.gob.ar}}
\affil{$^4$Instituto Nacional de Pesquisas Espaciais (INPE), Brazil}}
\affil{$^5$Universidad Tecnol{\'o}gica Nacional (UTN-FRM) y CONICET; Mendoza, Argentina; \email{icabello@mendoza-conicet.gob.ar}}
\affil{$^6$Universitat de Valencia, Val{\`{e}}ncia, Spain; \email{vdomingo@uv.es}}

\paperauthor{J.~Palacios}{judith.palacios@uah.es}{}{Universidad de Alcal{\'a}, UAH}{Departamento de F{\'{\i}}sica y Matem{\'a}ticas}{Alcal{\'a} de Henares}{Madrid}{28871}{Spain}
\paperauthor{S.~Vargas Dom{\'{\i}}inguez}{svargasd@unal.edu.co}{}{Universidad Nacional de Colombia}{Observatorio Astron\'omico Nacional}{Bogot{\'a}}{Bogot{\'a}}{111321}{Colombia}
\paperauthor{L.~A.~Balmaceda}{lbalmaceda@icate-conicet.gob.ar}{}{Author2 Institution}{Author2 Department}{City}{State/Province}{Postal Code}{Country}
\paperauthor{I.~Cabello}{icabello@mendoza-conicet.gob.ar}{}{Universidad Tecnol{\'o}gica Nacional and Consejo Nacional de Investigaciones Cient'f{\'{\i}}cas y T{\'e}cnicas}{Facultad Regional Mendoza}{Mendoza}{Mendoza}{5500}{Argentina}
\paperauthor{V.~Domingo}{vdomingo@uv.es}{}{Universitat de Vall{\`{e}}ncia}{GACE/IPL}{Paterna}{Valencia}{46980}{Spain}

\begin{abstract}

In this work we follow a series of papers on high-resolution observations of small-scale structures in the solar atmosphere \citep[][Cabello et al., in prep]{Balmaceda2009, Balmaceda2010, Vargas2011, Palacios2012, Domingo2012, Vargas2015}, combining several  multi-wavelength data series. These were acquired by both ground-based (SST) and space-borne (Hinode) instruments during the joint campaign of the Hinode Operation Program 14, in September 2007. Diffraction-limited SST data were taken in the G-band and G-cont, and were restored by the MFBD technique. Hinode instruments, on the other hand, provided multispectral data from SOT-FG in the CN band, and Mg~{\sc I} and Ca {\sc II}~lines, as well as from SOT-SP in the Fe~{\sc I} line. In this series of works we have thoroughly studied vortex flows and their statistical occurrences, horizontal velocity fields by means of Local Correlation Tracking (LCT), divergence and vorticity. Taking advantage of the high-cadence and high spatial resolution data, we have also studied bright point statistics and magnetic field intensification, highlighting the importance of the smallest-scale magnetic element observations.

\end{abstract}

\section{Introduction}

Photospheric flows, tracked by the granular motion imaged in the G or CN bands, are the manifestation of the convection that take place below the visible surface of the Sun. From the horizontal velocity fields obtained by following the constant motion of the granules, it is possible to derive maps of divergence and detect source or sink regions, depending on the horizontal flow divergence. Sources correspond mainly to granules, being the largest the exploding granules. On the other hand, sinks are always located in the intergranular lanes, and more specifically, in intergranular junctions. In such regions flows may exhibit a bathtub motion before converging into downflows. Coriolis forces are negligible in these small-scale swirls. These bathtub-like motions can be studied by different means \citep[see the extensive introduction in][]{Vargas2015}.

The study of small-scale vortex motions in the solar atmosphere is of great importance in the context of solar dynamo. Investigating their properties such as how are they distributed on the solar surface, their rate of occurrence and lifetimes can shed some light on the presence of a turbulent dynamo operating below the surface.

\section{Methods}

Data were mainly obtained during the Hinode HOP 14, but also HOP 72. Ground-based images from SST were restored with the Multi-Frame Blind Deconvolution \citep[MFBD, ][]{Lofdahl2002} and then postprocessed to remove other distortions. Data from Hinode were processed with the standard reduction scheme and p-mode removal.

In order to estimate the horizontal velocity fields, and consequently divergence and vorticity, we used the Local Correlation Tracking \citep[LCT,][]{November1988}. This technique also provides an estimate of vertical velocities, when direct methods to detect downflows (for instance, dopplershift measurements) are unavailable.

Bright points (BPs) observed in the G- and CN-bands or even in the Ca~{\sc II} line can be used as vortex flow tracers. In this case, we used methods to distinguish and classify BPs from granular fragments, based in segmentation and recognition pattern procedures. 
We mainly used and compared two methods: an automatic technique called MLT4, \citet{Bovelet2007}; and also a manual algorithm described in \citet{SanchezAlmeida2004}. We have also tracked the motion of small-scale magnetic structures via the center-of-gravity calculation using line-of-sight Mg {\sc I} magnetograms.

Using SOT-SP data we inferred magnetic flux densities, dopplershift and temperatures via inversion. We applied the LTE full-atmosphere inversion code LILIA \citep{SocasNavarro2001} and also used weak-field approximation for further comparison.

\section{Main results}

Here we summarize the main results from the series the aforementioned works. 

The multi-wavelength study (G-band, CN, Ca{\sc II}, Mg {I} magnetograms) of a small scale magnetic structure presented in \cite{Balmaceda2010} showed two clumps of magnetic elements rotating around a common center in a vortical motion, rotating almost 360$\deg$ for more than an hour and having associated downflows.  

After the analysis of that case study, we performed an intensive search of similar events in the whole field-of-view of SST data \citep{Vargas2011}. We found values of vortex occurrence of about 1.5$\cdot$10$^{-3}$~Mm$^{-2}\cdot$min$^{-1}$, and density values of 3$\cdot$10$^{-2}$~Mm$^{-2}$. The mean horizontal speeds is about 0.5 km$\cdot$s$^{-1}$, however, this mean value increases when vortices increase the circulation. The representative vortex radius is 250 km. The most common downflow speed (inferred from the divergence) is about 0.5 km$\cdot$s$^{-1}$.

Also, the properties of magnetic elements observed at different wavelengths, as in the CN and G bands and Mg {\sc I} magnetograms were also studied \citet[][Cabello et al., in prep]{Balmaceda2009}. The inferred typical diameters are 0.\arcsec27 for CN, and 0.\arcsec14 for G-band. We have to take into account the different diffraction limits for Hinode (CN) and SST (G-band). For the magnetic elements, the diameter were about 0.\arcsec44, observed in Hinode magnetograms. The usual characteristic lifetimes are from 5-10 minutes. 

Finally, in \citet{Vargas2015}, a deeper analysis of the case study focused on the flow dynamics and behaviour of the small magnetic elements, which present intensity variations, coalescence and disappearance as the main structure is observed to rotate. Significant downflows and episodes of magnetic field intensification are detected in the region where BPs are located.

This integral set of observations allowed us to determine some important characteristics of both magnetic elements and flows at small scales; as well as to study their behavior and evolution in short time scales as observed in the solar surface. 
Further similar studies based on both ground-based and space-borne data with extremely high resolution are fundamental to improve our current understanding of the role of the smallest scale features on the solar magnetism. 


\acknowledgements J.~P. acknowledges funding for UAH-travel grants for CSPM, and MINECO project no. AYA2013- 47735-P. I.~C. acknowledges CONICET postdoctoral grant and project MSUTIME0002218TC funded by UTN.
Hinode is a Japanese mission developed and launched by ISAS/JAXA, with NAOJ as domestic partner and NASA and STFC (UK) as international partners. It is operated by these agencies in co-operation with ESA and NSC (Norway). The Swedish 1-m Solar Telescope is operated by the Institute of Solar Physics of the Royal Swedish Academy of Sciences at the Spanish Observatorio del Roque de los Muchachos of the Instituto de Astrof{\'{\i}}sica de Canarias. We acknowledge projects ESP2006-13030-C06-04 and AYA2009-14105-C06.

\bibliography{cspm2015_palacios2.bib}  

\end{document}